\documentclass[12pt]{article}
\usepackage{epsfig,latexsym}
\date{\today}
\def\bm{\boldmath}
\def\be{\begin{equation}}
\def\ee{\end{equation}}
\def\bear{\be\begin{array}}
\def\bea{\begin{eqnarray}}
\def\eea{\end{eqnarray}}

\def\lsi{\raise0.3ex\hbox{$<$\kern-0.75em\raise-1.1ex\hbox{$\sim$}}}
\def\gsi{\raise0.3ex\hbox{$>$\kern-0.75em\raise-1.1ex\hbox{$\sim$}}}
\def\lsim{\mathop{\lsi}}
\def\gsim{\mathop{\gsi}}

 \def\Pom{{\bf I\!P}}

\def\Qb{\overline{Q}}

\newsavebox{\fmbox}
    
\title {\bf \bm
Shrinkage vs. anti-shrinkage of the diffraction cone
in the exclusive vector mesons production}

 \author{I.P.~Ivanov\thanks{E-mail: i.ivanov@fz-juelich.de}\\
 \makebox[8cm][c]{\normalsize Sobolev Institute of Mathematics, Novosibirsk, Russia}\\
 \makebox[8cm][c]{\normalsize IKP, Forschungszentrum J\"ulich, Germany}\\
 }

\begin{document}

\maketitle

 \vspace{-9cm}
 \makebox[\textwidth][r]{\large\bf FZJ-IKP(Th)-2003/XX}
 \vspace{7.5cm}

\abstract{
We investigate the energy behavior of the diffraction
cone in the exclusive vector meson production in diffractive DIS
within the $k_t$-factorization approach. In our calculations,
we make full use of fits to the unintegrated gluon structure functions
extracted recently from experimental data on $F_{2p}$.
Confirming early predictions, we observe that shrinkage
of the diffraction cone due to the slope of the Pomeron trajectory
is significantly compensated by the anti-shrinkage behavior
of the $\gamma \to V$ transition. In order to match 
recent ZEUS data on the energy behavior of the diffraction slope,
$\alpha^\prime_{eff}(J/\psi\mbox{, exp.}) = 0.115 \pm 0.018 (stat.) 
{}^{+0.008}_{-0.015}(syst.) \mbox{ GeV}^{-2}$,
we had to use an input value $\alpha^\prime_\Pom = 0.25$ GeV$^{-2}$.
We investigate the compensation effect in detail and give predictions
for $Q^2$-dependence of the rate of cone shrinkage
for different vector mesons.}

\section{Introduction}

The exclusive production of light and heavy vector mesons in diffractive DIS
\be
\gamma^{(*)} p \to Vp\label{gampvp}
\ee
is an ideal testing ground \cite{VMfirst,NNNVM,NNZ94} 
of the color-dipole approach, \cite{dipole1,dipole2,CTreviews},
and of the $k_t$-factorization approach,
\cite{VMtalks,disser,rhoscale}.
The two approaches are linked by the transverse Fourier-Bessel transform
from the color dipole size space to the momentum space: the former is based 
on the color dipole cross section, while the key ingredient in the latter approach
is the unintegrated gluon structure function ${\cal F}(x_g,\vec\kappa^2)$
of the proton. 
Recently, the unintegrated gluon density was
extracted from the experimental data on $F_{2p}$
and presented in a form of simple, ready-to-use parametrizations
for $x_g < 0.01$ and throughout the entire phenomenologically relevant
domain of gluon momenta $\vec\kappa$, \cite{IN2001}. 
These fits now allow one to
put many of previous qualitative predictions of the color dipole
or $k_t$-factorization approaches to quantitative grounds.
For example, they have already been successfully used in calculation
of (virtual) photoproduction of dijets off proton or nuclei
\cite{NSS,szczurek}. In the vector meson production,
these fits have helped understand the DGLAP factorization scale
in the case of $\rho$ meson production, \cite{rhoscale}, confirming
the earlier ideas of \cite{NNZ94}. It seems now timely
to reanalyze another issue that critically depends on the properties
of the gluon content of the proton, namely, the shrinkage
of the diffraction cone in the vector meson production.\\

In simple Regge-type models with an approximately 
linear Pomeron trajectory, the (momentum transfer) $t$-dependence of the 
intercept
\be
{d\sigma \over dt} \propto \left({W \over W_0}\right)^{4[\alpha_\Pom(t) -1]}\,,
\quad \alpha_\Pom(t) = \alpha_\Pom(0) + \alpha_\Pom^\prime\,t\,,
\ee
forces growth of the slope $b$ of the diffraction cone
\be
b(W) = b(W_0) + 4\alpha^\prime_\Pom \log\left({W \over W_0}\right)\,,
\ee 
the phenomenon called the shrinkage of the diffraction cone.
In QCD, the Regge limit is described by the BFKL equation \cite{BFKL}.
However, in the fixed coupling constant regime, the BFKL equation 
lacks any intrinsic dimensional scale 
and leads to the fixed value of the intercept.
Inclusion of the running coupling constant splits the fixed cut on the $j$ plane
into an infinite sequence of moving Regge poles \cite{Lipatov86},
each of them having its own non-zero $\alpha^\prime$.
Within the color dipole BFKL approach \cite{gBFKL},
the value of $\alpha^\prime_\Pom$ was obtained from the solution
of color dipole BFKL equation for the slope \cite{NZZ96}; its value was shown 
to be nonzero \cite{NZZ94alphaprime} 
and to depend both on the energy and the size of the color dipole. 
At very high energies, when the BFKL asymptotics
would fully develop and the saturation effects would not yet
come into play, the slope of the Pomeron trajectory would tend 
to a certain constant value independent of the dipole size.
This asymptotic value is governed by the gluon screening radius
and in \cite{NZZ96}, under certain initial conditions, 
it was estimated to be $\alpha^\prime_\Pom (asymp.) = 0.072$ GeV$^{-2}$. 
At HERA, however, we still reside in the subasymptotic region,
and the values of $\alpha^\prime_\Pom$ can be 
still $W$ and $Q^2$ dependent.

If we now return to the process (\ref{gampvp}),
we find that the shrinkage of the cone due to the non-zero
$\alpha^\prime_\Pom$ is only a part of the truth.
Thanks to the factorization properties,
the diffraction slope in the vector meson production
can be decomposed into the beam, target and exchange contributions.
A detailed analysis of each contribution performed in \cite{NNPZZ97} 
showed that the $\gamma \to V$ transition (beam) contribution
might possess substantial subasymptotic behavior. This energy behavior
was expected to be of the {\em anti-shrinkage type}, {\em i.e.}
the corresponding slope contribution was predicted to {\em decrease} with energy growth.
This partially compensates the shrinkage of the exchange contribution 
to the slope, and, therefore, leads to the effective rate
of the cone shrinkage observed in a vector meson $V$
production $\alpha_{eff}^\prime(V)$ smaller than $\alpha^\prime_\Pom$.

Quantifying this compensation requires the knowledge of the gluon density
in the soft region\footnote{To be accurate, one should understand the words 
``{\em soft gluon density}" only as 
``{\em the dipole-proton interaction in the soft region,
parametrized in terms of gluon density}".}.
At the time of publication \cite{NNPZZ97}, it was known rather poorly. 
Now, thanks to the fits obtained in \cite{IN2001}, we have 
a much better understanding of how the gluon density behaves
in the soft region. Therefore, it is possible now to reanalyze
the anti-shrinkage properties of $\gamma \to V$ transition 
and check whether this compensation effect is important.

This is done in the present work.
On the basis of the $k_t$-factorization approach, we calculate 
the differential cross section of the exclusive vector meson production
in diffractive DIS. We observe that the compensation effect is very important
numerically. As we show below, the input parameter 
$\alpha^\prime_\Pom = 0.25$ GeV$^{-2}$ leads to
the effective rate of the cone shrinkage in $J/\psi$ photoproduction
production $\alpha_{eff}^\prime \approx 0.12$ GeV$^{-2}$.  
This reduction implies that much care should be taken 
when the vector meson production results are interpreted as
a direct probe of the properties of the Pomeron.\\

The structure of this paper is following.
In Section 2 we briefly remind the basic formulae of the $k_t$-factorization
approach to the calculation of the vector meson production in diffractive DIS. 
By performing the small-$t$ expansion of a generic helicity-conserving 
amplitude, we qualitatively study different contributions to the diffraction 
slope and discuss the origin of the anti-shrinkage properties
of the $\gamma \to V$ transition. We then proceed to numerical
results, which are presented in Section 3. These results are discussed 
in Section 4, and Section 5 contains conclusions of this work.

\section{The $k_t$-factorization predictions for the diffraction slope:
a qualitative analysis}

\subsection{The basic amplitude}

The basic formulae for the vector meson production
within the $k_t$-factorization approach are well known
(see details in \cite{disser}).
We denote the quark and gluon loop transverse momenta 
and the momentum transfer 
by $\vec k$, $\vec \kappa$, and $\vec\Delta$, respectively
(here, vector sign denotes transverse vectors). 
The fraction of the photon lightcone momentum carried by the quark is denoted
by $z$, while the fractions of the proton light cone momentum carried
by the two gluons are $x_1$ and $x_2$. 
With this notation, the imaginary part of the amplitude 
of reaction (\ref{gampvp}) can be written in a compact form:
\be
Im {\cal A}= s{c_{V}\sqrt{4\pi\alpha_{em}}\over 4\pi^{2}}
\int {d^{2} \vec \kappa \over \vec\kappa^{4}}\alpha_{S}(q^2)
{\cal{F}}\left(x_1,x_2,\vec\kappa_1,\vec\kappa_2\right)
\int {dzd^2 \vec k \over z(1-z)}  \psi^*_V(z,\vec k)
\cdot I(\lambda_V,\lambda_\gamma)\, .
\label{f1}
\ee
where $\vec\kappa_{1,2} = \vec \kappa \pm {1 \over 2}\vec \Delta$.
The helicity-dependent integrands $I(\lambda_V,\lambda_\gamma)$ 
have form
\begin{eqnarray}
I^S(L,L) &=& 4 QM z^2 (1-z)^2
\left[ 1 + { (1-2z)^2\over 4z(1-z)} 
{2m \over M+2m}\right] \Phi_2\,;\nonumber\\[1mm]
I^S(T,T) &=& (\vec{e}\vec{V}^*)[m^2\Phi_2 + (\vec{k}\vec{\Phi}_1)] 
+ (1-2z)^2(\vec{k}\vec{V}^*)(\vec{e}\vec{\Phi}_1){M \over M+2m}\nonumber\\
&&- (\vec{e}\vec{k})(\vec{V}^*\vec{\Phi}_1) + 
{2m \over M+2m}(\vec{k}\vec{e})(\vec{k}\vec{V}^*)\Phi_2\,;\nonumber\\[1mm]
I^S(L,T) &=& 2Mz(1-z)(1-2z)(\vec{e}\vec{\Phi}_1)
\left[ 1 + { (1-2z)^2\over 4z(1-z)} {2m \over M+2m}\right]\nonumber\\[1mm]
&&- {Mm\over M+2m}(1-2z)(\vec{e}\vec{k})\Phi_2\,;\nonumber\\[1mm]
I^S(T,L) &=& -2Qz(1-z)(1-2z)(\vec{V}^*\vec{k}){M \over M+2m}\Phi_2\,,\label{f8}
\end{eqnarray}
where
\begin{eqnarray}
\Phi_2& =& -{1 \over (\vec{r}+\vec\kappa)^2 + \varepsilon^2} -{1 \over
(\vec{r}-\vec\kappa)^2 + \varepsilon^2} + {1 \over (\vec{r} + \vec\Delta/2)^2 +
\varepsilon^2} + {1 \over (\vec{r} - \vec\Delta/2)^2 + 
\varepsilon^2}\nonumber\,;\\[5mm]
\vec{\Phi}_1 &=& -{\vec{r} + \vec\kappa \over (\vec{r}+\vec\kappa)^2 + \varepsilon^2}
-{\vec{r} - \vec\kappa \over (\vec{r}-\vec\kappa)^2 + \varepsilon^2}
+ {\vec{r} + \vec\Delta/2 \over (\vec{r} + \vec\Delta/2)^2 + \varepsilon^2}
+ {\vec{r} - \vec\Delta/2 \over (\vec{r} - \vec\Delta/2)^2 + 
\varepsilon^2}\nonumber\,,
\end{eqnarray}
with $\vec r \equiv \vec k - (1-2z)\vec\Delta/2$ and 
$\varepsilon^2 = z(1-z)Q^2 + m_q^2$. Finally, the strong coupling
constant is taken at 
$q^2 \equiv \mbox{ max}[\varepsilon^2 + \vec k^2, \vec\kappa^2]$.

In the absence of $\vec\Delta-\vec\kappa$ correlations,
and for a very asymmetric gluon pair, the off-forward gluon structure function 
${\cal{F}}\left(x_1,x_2,\vec\kappa_1,\vec\kappa_2\right)$ 
that enters (\ref{f1}) can be approximately related to the forward 
gluon density ${\cal F}(x_g,\vec \kappa)$ via
$$
{\cal F}(x_1,x_2\ll x_1,\vec\kappa_1,\vec\kappa_2) 
\approx {\cal F}(x_g,\vec\kappa) \cdot 
\exp\left(- {1 \over 2}b_{3\Pom}\vec\Delta^2\right)\,.
$$
Here $x_g \approx 0.41 x_1$; the coefficient $0.41$ 
is just a convenient representation
of the off-forward to forward gluon structure function relation
found in \cite{shuvaev}.
Numerical parametrizations of the forward unintegrated gluon density 
 ${\cal F}(x_g,\vec \kappa)$ for any practical values of $x_g$ and 
$\vec \kappa^2$ can be found in \cite{IN2001}.
The slope $b_{3\Pom}$ contains contributions from the proton
impact factor as well as from the Pomeron exchange;
experimentally, it can be accessed in the
high-mass elastic diffraction.

The vector meson wave function $\psi_V(z,\vec k)$ describes 
the projection of the $q\bar q$ pair onto the physical vector meson.
It is normalized so that the forward value of the vector meson
formfactor is unity, and the free parameters are chosen 
to reproduce the experimentally observed value of the
vector meson electronic decay width $\Gamma(V \to e^+e^-)$.
In what concerns the shape of the radial wave function,
we followed a pragmatic strategy. We took a simple
Ansatz for the wave function, namely, the oscillator type wave function
and performed all calculations with it. In order
to control the level of uncertainty,
introduced by the particular choice of the wave function,
we redid the calculations with another wave function Ansatz, namely,
the Coulomb wave function, and compared the results.
Since these two wave functions represent the two extremes (very compact
and very broad wave functions that still lead to the same value of 
the electronic decay width), the difference observed should give
a reliable estimate of the uncertainty.
This difference is typically given by factor of 1.5 for absolute values
of the cross sections, while in the observables that involve
ratios of the cross sections (such as slopes, intercepts, etc.) 
this uncertainty is reduced. Details can be found in \cite{disser,INS2003}.

Note also that when deriving (\ref{f8}), we treated the vector mesons
as $1S$ wave states and used the pure $S$-wave spinorial structure 
${\cal S}^\mu$ instead of $\gamma^\mu$, \cite{IN99}.

\subsection{The contributions to the diffraction slope}

The main feature of the $|t| \equiv \vec\Delta^2$-dependence 
of the cross sections is a pronounced forward diffraction
cone, which can be, at very small $\vec\Delta^2$, parametrized
by a single slope parameter $b$ (see Section 3.1 for a detailed discussion
on the definition of the slope). 
Let us understand the various contributions to the slope by
studying the small-$t$ expansion of a typical amplitude.
Due to factorization properties,
one can approximately decompose the slope to the target, exchange 
and the projectile contribution, \cite{NNPZZ97}:
\be
b = b_{p\to p}  + b_{\Pom} + b_{\gamma^*\to V}\,.\label{slope-decomp}
\ee
In principle, the presence of the helicity-flip amplitudes 
represents yet another source of the $t$-dependence.
Since the single helicity-flip amplitudes are proportional to $\sqrt{|t|}$,
one can introduce correction to the slope as
$$
d\sigma/dt_{non-flip} + d\sigma/dt_{flip} = A(t) + B\cdot|t| \to
A(t)\cdot e^{-b_{flip} |t|}\,.
$$
However, since the helicity-flip amplitudes are small, this correction 
never exceeds 2\% of the value of the slope.

The first term in (\ref{slope-decomp}) is an intrinsically 
soft quantity and is related to the wave function
of the proton. It is not calculable within pQCD, but its magnitude can be 
estimated from the proton charge radius. In our calculations, 
we introduced the following elastic formfactor to the amplitudes
\be
F(\vec\Delta^2) = {1 \over \left(1+\vec\Delta^2/\Lambda^2\right)^2}\,;
\quad\mbox{with}\ \Lambda = 1\mbox{ GeV},
\ee
which leads to the proton impact factor contribution to the slope
\be 
b_{p\to p} = 4 \mbox{ GeV}^{-2}\,.\label{slopefirst}
\ee
The second contribution in (\ref{slope-decomp}) arises from the $t$-dependence 
of the Pomeron intercept. In our calculations, it was parametrized as
\be
b_{\Pom} = \alpha^\prime_\Pom \log\left({x_0 \over x_g}\right)\,;\quad
\alpha^\prime_\Pom = 0.25 \mbox{ GeV}^{-2}\,,
\quad x_0 = 3.4\cdot 10^{-4}\,.\label{slopesecond}
\ee
This dependence was ascribed both to hard and soft components of the
unintegrated gluon structure function, see \cite{IN2001}; more sophisticated
parametrizations can be put forth as well.
This parametrization was obtained by the requirement
that we describe well the recent ZEUS data on the shrinkage 
of $J/\psi$ photoproduction \cite{ZEUSjpsiphoto}.
We underline that, due to the {\em compensation effect} 
to be explained in a minute,
this $\alpha^\prime_\Pom = 0.25 \mbox{ GeV}^{-2}$ reduced to 
$\alpha^\prime_{eff} \approx 0.12 \div 0.13$ GeV$^{-2}$ 
in the differential cross section, in agreement with the experimental data.

Finally, the third term in (\ref{slope-decomp}) originates 
from the photon to vector meson transition amplitude. 
This contribution possesses both $Q^2$ and energy dependence,
which can be understood as follows.

We start with the amplitude $L\to L$ and consider function 
$\Phi_2$ at large enough values of $\varepsilon^2$ so that $\vec k^2$
can be safely neglected. Expand it at small $\vec\Delta^2$ and average over
all possible directions of $\vec\Delta$. The result reads
\be
\Phi_2 \approx {2\vec\kappa^2 \over \varepsilon^2(\vec\kappa^2 + \varepsilon^2)}
- {\vec\Delta^2 \over (\vec\kappa^2 + \varepsilon^2)^3}
\left[{1 \over 2}(\varepsilon^2 - \vec\kappa^2) + 4[z^2+(1-z)^2]\vec\kappa^2
\left(1 + {3 \vec\kappa^2\over 4 \varepsilon^2} + {\vec\kappa^4\over4 \varepsilon^4}\right) \right]\,.
\label{slope3}
\ee
To the leading log $Q^2$, $\vec\kappa^2 \ll \varepsilon^2$, and one has
$$
\Phi_2 \approx {2\vec\kappa^2 \over \varepsilon^4} 
- 2 (1-2z)^2 {\vec\Delta^2 \vec\kappa^2\over \varepsilon^6}
- {\vec\Delta^2 \over 2\varepsilon^4}\,.
$$
After performing the relevant integration, one obtains that the amplitude 
$L\to L$ is proportional to
$$
{2 \over \Qb_L^4} G(x_g,\Qb_L^2) - {2 \over \Qb_L^4} G(x_g,\Qb_L^2)
\cdot \eta_L {\vec\Delta^2 \over \Qb_L^2} - {\vec\Delta^2 \over 2\Qb_L^4 } 
{{\cal F}(x_g,\mu^2) \over \mu^2}\,.
$$
Here $G(x_g,\Qb_L^2)$ is the conventional gluon density,
$\eta_L = \langle (1-2z)^2\rangle_L$ and $\mu^2$ 
is an appropriately defined soft scale. 
The relevant hard scale $\Qb^2_L$, which is linked to the scanning
radius of the color dipole approach $r_s^2 \leftrightarrow 1/\Qb^2_L$,
was investigated in detail in \cite{rhoscale}.
Thus, one obtains the following contributions to the slope
\be
b_{\gamma\to V} = {2 \eta \over \Qb_L^2} + {1 \over \mu^2} {{\cal F}(x_g,\mu^2)
\over G(x_g,\Qb_L^2)}\,.\label{slopethird}
\ee
The first term in (\ref{slopethird}) is a perturbative contribution.
At $Q^2=0$, it should be of the order of several GeV$^{-2}$, 
but it quickly falls off with the $Q^2$ growth. The second contribution
is a predominantly soft quantity, its dependence on $Q^2$ is weak.

The $T\to T$ amplitude can be analyzed in a similar way, and the 
qualitative results of this analysis remain the same as for the longitudinal
case. This is why we use the above qualitative results 
even when discussing the properties of the transverse amplitudes.

\subsection{The compensation effect: sources of the cone anti-shrikage}
\label{anti}

The total slope of the diffraction cone is given by the sum of all three
contributions, (\ref{slope-decomp}). Since the target contribution to the slope
is taken constant, the energy behavior of the slope originates only from
$b_\Pom$ and $b_{\gamma \to V}$.

The exchange contribution to the diffraction slope, $b_\Pom$, logarithmically
grows with energy and leads to the well-known shrinkage of the diffraction cone.
In a simple Regge-type models, the properties of the Pomeron coupling
to hadrons are assumed to be energy independent, and this contribution
is the only source of the energy dependence of the diffraction cone.
In a more elaborate models, such as the $k_t$-factorization approach,
the beam contribution to the diffraction slope, $b_{\gamma \to V}$,
also possesses the energy dependence.

On the basis of the above qualitative analysis, one can
identify two dinstinct sources of this energy behavior.
First, even within the leading $\log Q^2$, the second term in (\ref{slopethird})
depends on energy. Indeed, both ${\cal F}(x_g,\mu^2)$ and
$G(x_g,\Qb_L^2)$ rise with energy, but the exponents
of their rise  are different, see \cite{IN2001}. 
For $\Qb_L^2 \gsim 1$ GeV$^2$,
the integrated glue taken at $\Qb_L^2$ rises with energy faster than
the differential glue at the soft scale $\mu^2$, and this contribution
to the slope will {\em decrease} with energy rise.
This source of the energy behavior of $b_{\gamma \to V}$ was also
discussed in \cite{NNPZZ97}.

\begin{figure}[!htb]
   \centering
   \epsfig{file=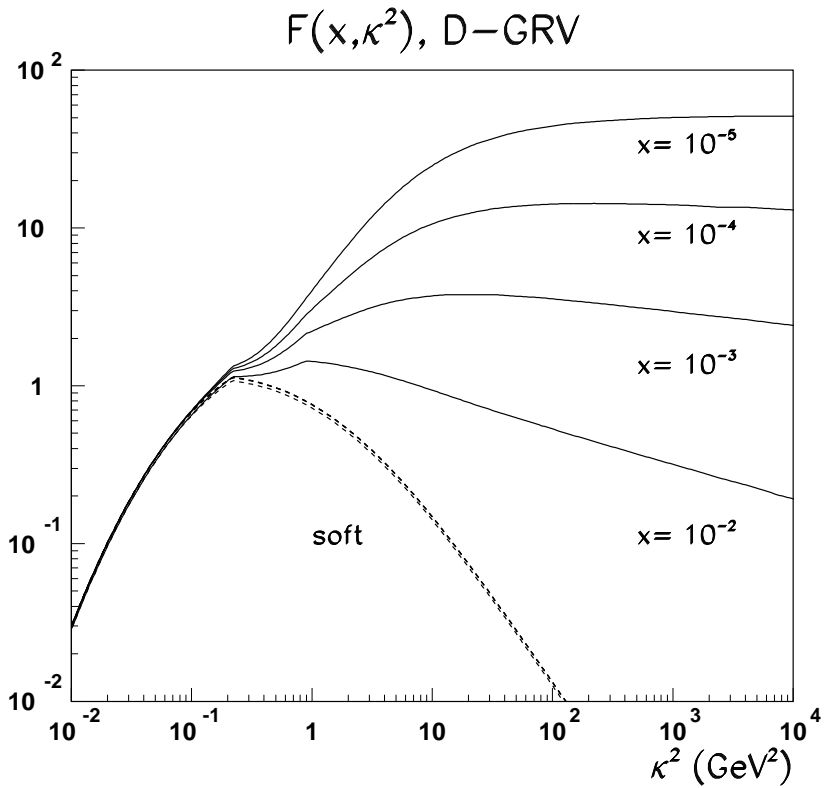,width=120mm}
   \caption{The $\vec\kappa$-dependence of the unintegrated gluon density
   ${\cal F}(x_g,\vec\kappa^2)$ (fit D-GRV) for several values of $x_g$.
   The solid and dashed curves correspond, respectively, 
   to the full gluon density and to the soft contribution only.}
   \label{d-grv}
\end{figure}

The second source of the $b_{\gamma \to V}$ energy behavior
appears, if we consider (\ref{slope3}) beyond the leading log $Q^2$.
In this case denominators will contain $\Qb_L^2 + \vec\kappa^2$
instead of just $\Qb_L^2$.  Due to specific properties of the 
unintegrated gluon density, the typical values
of $\vec\kappa^2$ grow with energy even at fixed $\Qb_L^2$.
This is clear from Fig.~\ref{d-grv}, taken from \cite{IN2001}, where
we showed, in a single plot, how the unintegrated gluon density 
${\cal F}(x_g,\vec\kappa^2)$ changes with energy (or $1/x_g$) growth. 
One sees that the relative weight of the large $\vec\kappa^2$ region
increases as we move from $x_g = 10^{-2}$ to $x_g = 10^{-4}$.
This proves that $\Qb_L^2 + \vec\kappa^2$ increases ---
and the contribution (\ref{slopethird}) to the slope 
again {\em decreases} --- with the energy rise. 
This effect mostly relies on specific, yet unavoidable, properties of the 
gluon density.

Both sources of the energy behavior of the beam contribution to the slope
are of the {\em anti-shrinkage type}. They lead to
the partial compensation of the diffraction cone shrinkage, 
and eventually result in $\alpha^\prime_{eff} < \alpha^\prime_\Pom$.

It is interesting to note that one can, in principle, 
disentangle these two sources of the compensation effect.
What one needs is the study of the $k_t$-factorization prediction 
for the ultrahigh energy behavior of $\alpha^\prime_{eff}$.
Note that the second contribution to the compensation effect
works at full strength only at $x_g > 10^{-4}$. At smaller $x_g$,
{\em i.e.} at higher energies,
the $\vec\kappa$-shape of the unintegrated gluon density
practically does not change, and the growth of average values of $\vec\kappa^2$
with energy stops. Thus, in this region, the compensation
effect should be only due to the first mechanism.

\section{Numerical results}

\subsection{Definitions of the slope}

\begin{figure}[!htb]
   \centering
   \epsfig{file=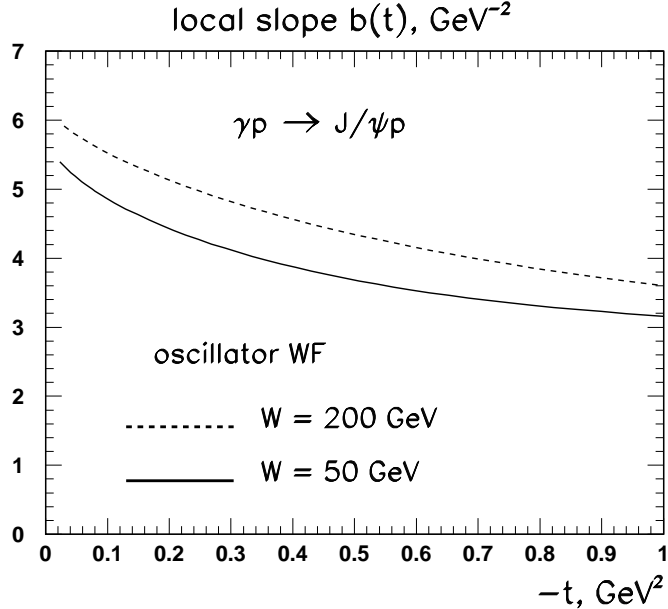,width=100mm}
   \caption{The momentum transfer dependence of the local slope 
(\ref{deflocal}) for the photoproduction of $J/\psi$ meson.
The solid and dashed curves correspond to total energies of the $\gamma p$
collision equal to 50 GeV and 200 GeV, respectively.}
   \label{localslope-jpsi}
\end{figure}

Before comparing the results of the $k_t$-factorization predictions
of the diffraction slope with the experimental data,
we would like to discuss the definition of the slope itself.
The literal definition of the {\em local slope}
of the diffraction cone as a logarithmic defivative of the differential
cross section is
\be 
b(t) = - {d \over d|t|} \log\left({{d\sigma \over d|t|}}\right)\,.
\label{deflocal}
\ee
Since the differential cross section flattens as $|t|$
increases, the value of the local slope will decrease with $|t|$ growth.
In Fig.~\ref{localslope-jpsi}, we show the $k_t$-factorization calculation of the local
slope for the $J/\psi$ photoproduction within the region 
$|t| \le 1$ GeV$^2$.

The values of the diffraction slope obtained in the experiment
are often the results of exponential law fits to the measured differential 
cross section performed within a certain $t$ interval. 
This procedure approximately corresponds to 
a {\em finite-difference slope} defined as
\be 
b(t_1,t_2) =  {1 \over  |t_2| - |t_1|}\left(\log{d\sigma \over d|t|}\Big|_{t_1}
- \log{d\sigma \over d|t|}\Big|_{t_2}\right)\,,
\label{deffinite}
\ee
where $t_1$ and $t_2$ define the region of the experimental fit.
When comparing our predictions with such data, we will use
precisely this definition of the slope.

In literature, other definitions of the slope can be encountered, such as
${1 \over \sigma} {d\sigma \over dt}|_{t=0}$ or $1/\langle |t|\rangle$.
If the differential cross section were a pure exponential law,
all these definitions would lead to the same values.
However, in a realistic situation, the offset among them
can reach $\sim 1\div 2$ GeV$^{-2}$. This should be kept in mind
when comparing different results of the slope.

\subsection{The energy dependence of the slope parameter}

We start our analysis with the $J/\psi$ photoproduction.
The local slope parameter, predicted from the $k_t$-factorization
approach, has already been shown in Fig.~\ref{localslope-jpsi} 
within the region $|t| \le 1$ GeV$^2$ for two values of the $W_{\gamma p}$. 
One observes a steady growth of the slope as the energy increases,
which leads to the shrinkage of the diffraction cone.
Motivated by the Regge-model considerations, this growth is usually
described by the following law:
\be
b(W) = b(W_0) + 4\alpha_{eff}^\prime \log\left({W \over W_0}\right)\,.
\ee
The quantity $\alpha_{eff}^\prime$, which we will call the 
{\em rate of the diffraction cone shrinkage}, 
is often assumed to be equal to the slope of the Pomeron trajectory
$\alpha_\Pom^\prime$. However, as we argued in the previous section,
there are grounds to expect that these two will 
differ from each other.

\begin{figure}[!htb]
   \centering
   \epsfig{file=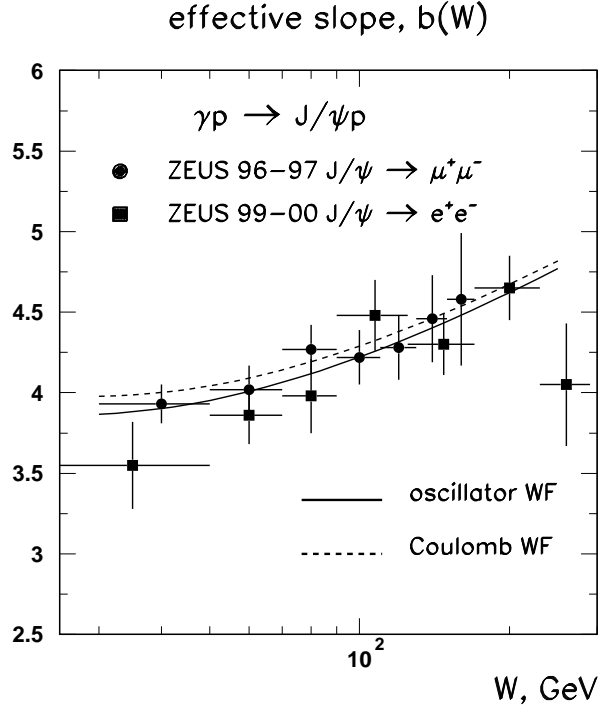,width=100mm}
   \caption{The energy dependence of the diffraction slope in $J/\psi$
photoproduction. Experimental data points are from ZEUS \cite{ZEUSjpsiphoto}.}
   \label{slope-jpsi-wdep}
\end{figure}

The rate of the shrinkage $\alpha_{eff}^\prime$ can be analyzed quantitatively
on a plot of $b$ as a function of energy $W$.
As can be seen from Fig.~\ref{localslope-jpsi}, 
the rate of the cone shrinkage is roughly independent
of the value of $|t|$, therefore one can expect
that different definition of the slope will still produce similar
$\alpha_{eff}^\prime$. In Fig.~\ref{slope-jpsi-wdep} we plotted
the $k_t$-factorization calculations of the energy dependence
of the diffraction slope $b(W)$ in the $J/\psi$ photoproduction.
In order to have control on the level
of uncertainty introduced by the wave function Ansatz, we calculated $b(W)$ 
for both the oscillator and the Coulomb wave functions.

The same Figure contains also the 
recent experimental data from ZEUS \cite{ZEUSjpsiphoto}.
Keeping in mind the discussion of the previous subsection,
we attempted in our calculations to match the experimental 
procedure of the diffraction slope evaluation.
The $k_t$-factorization values of the slopes 
were calculated according to (\ref{deffinite}) with 
$|t_1| = 0.1$ GeV$^2$ and $|t_2| = 0.7$ GeV$^2$. 
As we mentioned above, it is precisely these $J/\psi$ photoproduction 
experimental data that we use to fix the two free parameters 
of the diffraction slope (\ref{slopesecond}). 
Therefore the agreement between
the curves and the data points is nothing else but just the double-check
of the consistency of our calculations.

As can be seen from this Figure, the $k_t$-factorization calculations
do not predict $b(W)$ to be strictly linear function of $\log(W)$,
as at the low energy end of the plot the curves flatten out.
This is a consequence of the fact that the compensation effect
discussed above can be well energy dependent (see also next subsection).
However, at $W \gsim 70$ GeV, the approximate 
linearity is indeed observed.
Note also that $J/\psi$ photoproduction at low energies, 
$W \lsim 25$ GeV, corresponds to $x_g \gsim 0.01$. 
This is the very edge of the $x_g$ region
the gluon density fits were devised for. 
We prefer to eliminate potential problems with applicability of the 
gluon density parametrizations and focus on region $W \ge 50$ GeV.
If we now evaluate the shrinkage of the cone between $W = 50$ GeV
and $W = 250$ GeV, we find the $k_t$-factorization values
\be
\alpha^\prime_{eff} \approx \left\{ 
\begin{array}{ll}
0.127\mbox{ GeV}^{-2} & \mbox{for the oscillator wave function}\,,\\
0.121\mbox{ GeV}^{-2} & \mbox{for the Coulomb wave function}\,,
\end{array}
\right.\label{ktresults}
\ee
which are, of course, in agreement with the experimentally measured value
\cite{ZEUSjpsiphoto}
\be
\alpha^\prime_{eff}(\mbox{exp.}) = 0.115 \pm 0.018 (stat.) 
{}^{+0.008}_{-0.015}(syst.) \mbox{ GeV}^{-2}\,.
\ee
The key observation here is that the $\alpha^\prime_{eff}$ value predicted
by the $k_t$-factorization is sigfinicantly less than the input
parameter $\alpha^\prime_\Pom = 0.25$ GeV$^{-2}$.

\begin{figure}[!htb]
   \centering
   \epsfig{file=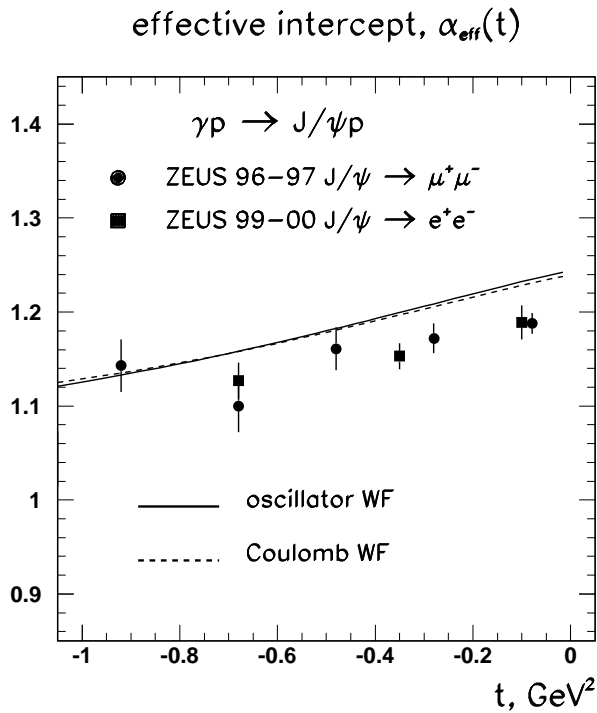,width=100mm}
   \caption{The $t$-dependence of the effective intercept 
$\alpha_{eff}(t)$ calculated from the energy growth of the differential
cross section of the $J/\psi$ photoptoduction between points $W=50$ GeV
and $W=250$ GeV. Experimental data points are from ZEUS, \cite{ZEUSjpsiphoto}.}
   \label{alpha-jpsi-tdep}
\end{figure}

Another look at how the diffraction cone behaves with the energy growth
is given by the energy rise of the differential cross section itself.
Introducing the effective intercept for the differential cross section
\be
{d\sigma\over dt}\Big|_W = {d\sigma\over dt}\Big|_{W_0} 
\cdot \left({W \over W_0}\right)^{4[\alpha_{eff}(t)-1]}\,,
\ee
one can study how this intercept changes with $t$. 
The results of this study,
together with ZEUS experimental data \cite{ZEUSjpsiphoto}, are shown
in Fig.~\ref{alpha-jpsi-tdep}. Although the experimental data were
available down to $|t| = 1.34$ GeV$^2$, we limited ourselves only 
to $|t|< 1$ GeV$^2$ region. Within this region, the $k_t$-factorization
calculations based on either of the two wave functions are in a qualitative
agreement with the data. Besides, one sees that the shape of the vector meson
wave function has rather minor effect on the diffraction cone shrinkage,
and, in the subsequent plots we will give only results based on the
oscillator wave function.

\subsection{The $Q^2$ and $W$ behavior of the compensation effect}

It is known experimentally, and it follows from our analysis 
in Section 2 as well,
that the values of the diffraction slope $b$ are sensitive to the 
virtuality $Q^2$ and to the mass of the vector meson produced $m_V$.
It is, therefore, interesting to check how the values of $Q^2$ and $m_V$
will affect the rate of the diffraction cone shrinkage $\alpha^\prime_{eff}$.

\begin{figure}[!htb]
   \centering
   \epsfig{file=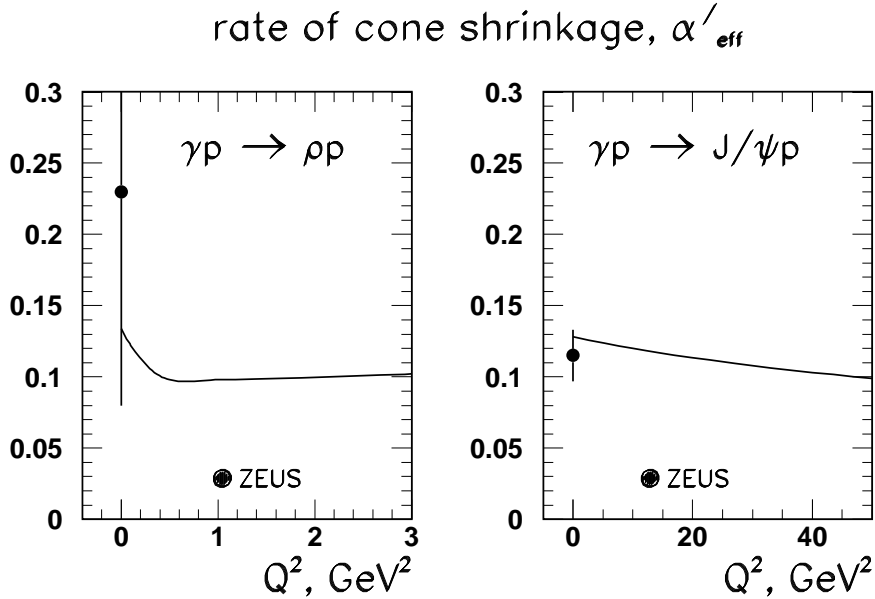,width=140mm}
   \caption{The effective rate of the shrinkage of the diffractive cone
 $\alpha^\prime_{eff}$ for $\rho$ meson and $J/\psi$ meson
production as functions of $Q^2$. The $k_t$-factorization results
are calculated between $W=50$ GeV and $W=250$ GeV, and for the oscillator wave function
only. The photoproduction data points are from ZEUS, \cite{ZEUSrhophoto,ZEUSjpsiphoto}.}
   \label{shrinkage}
\end{figure}

In Fig.~\ref{shrinkage} we show the $k_t$-factorization predictions
of the $Q^2$ behavior of $\alpha^\prime_{eff}(\rho)$ 
and  $\alpha^\prime_{eff}(J/\psi)$.
We also show the two ZEUS data points available 
\cite{ZEUSrhophoto,ZEUSjpsiphoto}, 
both of which correspond to the photoproduction limit.
What one sees on both plots is a slight decrease of $\alpha^\prime_{eff}$
from $\approx 0.13$ GeV$^{-2}$ down to  $\approx 0.10$ GeV$^{-2}$,
which takes place at typical $Q^2 \sim m_V^2$.
This rather weak sensitivity of $\alpha^\prime_{eff}$ 
to the values of $Q^2 + m_V^2$
is in accordance with \cite{NZZ96}, where, too, 
the value of $\alpha^\prime_{eff}$ was found to depend mostly on energy
but not on the scanning radius, which is related to the virtuality,
or to be more precise, to the hard scale $\Qb^2$.

As mentioned in the previous section, the ultrahigh energy behavior
of $\alpha^\prime_{eff}$ can help disentangle the two sources 
of the compensation effect. If $\gamma p$ collision energy grows 
to high enough values, such that $x_g \lsim 10^{-4}$ (which corresponds,
for $J/\psi$ photoproduction, to $W \gsim 300$ GeV), then the second source
of the anti-shrinkage behavior discussed in Sect.~\ref{anti} 
should get suppressed. As a result, what one probes in this region
is only the first contribution to the compensation effect. The overall
compensation should, therefore, weaken.

Numerical study shows that at $W \gg 100$ GeV the effective rate of the
diffraction cone shrinkage $\alpha^\prime_{eff}$ significantly rises,
in accordance with expectations. At $W = 1000$ GeV and $W = 4000$ GeV, 
$\alpha^\prime_{eff}$ rises up to $0.19$ GeV$^{-2}$ and $0.22$ GeV$^{-2}$,
respectively, which should be compared to (\ref{ktresults}) at HERA energies.
This is, first, in agreement with our expectation that the compensation effect
should decrease with energy growth. Second, this results can be treated
as an evidence that, at HERA energies, both sources of the anti-shrinkage effect are 
of comparable importance. 

We would like to stress that this analysis is just a self-contained investigation
of the energy behavior of the compensation effect,
which is the property the $k_t$-factorization approach.
We {\em do not} attempt to predict the energy behavior of $\alpha^\prime_{eff}$.
Such a prediction might be feasible only with a reliable information on the energy behavior
of the input parameter $\alpha^\prime_{\Pom}$. In our calculations,
we used constant value of $\alpha^\prime_{\Pom} = 0.25$ GeV$^{-2}$ for all energies,
which might be very inaccurate beyond the HERA energy range.

\section{Discussion}

\subsection{Is this the Pomeron trajectory?}

The experimental results presented in Fig.~\ref{alpha-jpsi-tdep}
were explicitly interpreted in \cite{ZEUSjpsiphoto} as a measurement
of the Pomeron trajectory. Here, we argue that this interpretation
is, at least, not that straightforward.

The most direct argument comes from the value of $\alpha^\prime_{eff}$.
The numerical calculations show that the rate of the 
diffraction cone shrinkage $\alpha^\prime_{eff}$ predicted
by the $k_t$-factorization approach is sigfinicantly less than the input
parameter $\alpha^\prime_\Pom$, which quantifies
the $t$-dependence of the intercept of the underlying Pomeron exchange.
Therefore, there exists a mechanism of compensation of the cone shrinkage.
The qualitative analysis of Section 2 suggests that the
energy decreasing contribution (\ref{slopethird}) 
of the $\gamma \to V$ transition to the overall diffraction slope
is in charge of this compensation effect.
In Sect.~\ref{anti} we pointed out two sources of this anti-shrinkage behavior,
and the subsequent analysis gave evidences that both of them were equally
important.

In order to make sure that there is no other source
for anti-shrinkage but the $\gamma \to V$ transition,
we made a double-check and switched off the $\vec\Delta$-dependence
of integrands $I^S(\lambda_V,\lambda_\gamma)$ in (\ref{f8}).
This made $\alpha^\prime_{eff}$ jump up 
to $\approx 0.24$ GeV$^{-2}$, which is very close
to the input value $\alpha^\prime_\Pom = 0.25$ GeV$^{-2}$.
This proves that it is precisely the  $\gamma \to V$ transition 
vertex that causes such a strong reduction 
of the rate of the diffraction cone shrinkage in vector meson production.

Thus, we are led to a conclusion that, first, the experimentally  measured
values of $\alpha^\prime_{eff}$ should not be interpreted directly as
the slope of the Pomeron trajectory. Second, as we checked,
$\alpha^\prime_{eff}$--$\alpha^\prime_\Pom$ relation 
is rather robust within the $k_t$-factorization scheme,
$$
\alpha^\prime_{eff} \approx \alpha^\prime_\Pom - 0.13 \mbox{ GeV}^{-2}\,,
$$
and the experimental data shown in Fig.~\ref{alpha-jpsi-tdep} 
might be in fact an evidence that the true value 
of the slope of the Pomeron trajectory, as measured in the $J/\psi$ photoproduction,
is around $0.25$ GeV$^{-2}$.

It is interesting to note that this value is close to what
is usually believed to be the soft\footnote{Ironically, 
several competing theoretical approaches
agree on this, see also \cite{DL98,jenk,FMS01}.} 
$\alpha^\prime$, see \cite{NZZ96}.
One can also quote a similar value measured in
the elastic hadronic interaction at high energies 
\cite{hadronic,corrections}, but, in its own turn, 
the single Pomeron exchange in these reactions
receive sizable absorption corrections \cite{corrections} 
and, therefore, these results 
should not be interpreted straightforwardly as well. 
In any case, due to the strong compensation effect,
we think that it would be premature to conclude that the measured value
of $\alpha^\prime_{eff}$ in $J/\psi$ photoproduction
is inconsistent with the soft Pomeron.

\subsection{Comments on other works}

We would like to mention that our approach to $\alpha^\prime_{eff}$
is different from approaches of \cite{jenk} and \cite{FMS01}.
In \cite{jenk} the differential cross section of vector meson production
was fitted to the experimental data with no attempt to
disentangle the real physics that leads to this form of the cross section.
An explicity non-linear Pomeron trajectory 
was introduced, which, in constrast to our predictions, 
leads to a strong $t$-dependence of $\alpha^\prime$.
Although in this model $\alpha^\prime(t=0) = 0.25$ GeV$^{-2}$,
it becomes twice smaller at $|t| = 12m_\pi^2 \approx 0.24$ GeV$^2$,
which allows the authors of \cite{jenk} to describe reasonably well the
ZEUS data on shrinkage in $J/\psi$ photoproduction, \cite{ZEUSjpsiphoto}.
It must be pointed out that this agreement is not surprising, since
the fits to the cross section were derived precisely from the $J/\psi$ 
photoproduction data. 
By the construction of their model, the authors of \cite{jenk}
directly relate the observed shrinkage in vector mesons photoproduction
to the Pomeron properties, $\alpha^\prime_{eff} = \alpha^\prime_{\Pom}$.
At the end, authors state that they ``have reached a deeper understanding
of the properties of dipole Pomeron", however, in the light
of the present paper's results, such statement looks rather questionable.

The authors of \cite{FMS01} worked in the color dipole formalism
and, as can be expected from the early work \cite{NNPZZ97}, they should
have observed the compensation effect studied in the present paper.
When discussing $\alpha^\prime$, they indeed mention a similar effect,
but they estimate it to be very inessential numerically, contrary
to the claim of our work. The origin of this discrepancy lies in the fact that
the authors of \cite{FMS01} overlooked the main source of the energy dependence
of the slope of $\gamma\to V$ transition and took only inessential
part into account. Namely, in their phenomenological 
parametrization of the slope, the authors of \cite{FMS01} included 
only the first term in (\ref{slopethird}) 
(this is the $\langle b^2\rangle$ term in the notation of \cite{FMS01}).
The authors of \cite{FMS01} neither took into account the second term
in (\ref{slopethird}), nor went beyond leading log $Q^2$ approximation 
by accounting for $\vec\kappa^2$ in this term, and, as a result,
missed the sizable anti-shrinkage effect discussed here. Therefore, the conclusion 
$\alpha^\prime_{eff} \approx \alpha^\prime_{\Pom}$ of \cite{FMS01}
was misleading. 

\section{Conclusions}

In this work we investigated, at the quantitative level,
the energy behavior of the diffraction cone in the exclusive
production of vector mesons. The work was conducted
 in the $k_t$-factorization scheme, closely related to the
familiar color dipole formalism, and was based on
recent fits to the unintegrated gluon density obtained in \cite{IN2001}.
The fact that we do not devise models for the gluon content but
instead heavily rely on the high-precision experimental data 
lends certain credence to the whole calculation.

Confirming the results of early work
\cite{NNPZZ97}, we observed that the shrinkage of the diffraction cone,
that could be expected from the Pomeron properties, is partially compensated
by the anti-shrinkage behavior of the $\gamma \to V$
transition vertex. We pointed out two sources of this compensation effects,
both of equal importance, and observed that this 
reduction of the cone shrinkage is very significant numerically.
In order to reproduce the experimentally observed value
of $\alpha^\prime_{eff}(J/\psi) \approx 0.115$ GeV$^{-2}$,
we had to take $\alpha^\prime_\Pom = 0.25$ GeV$^{-2}$ as an input parameter.
This value turned out to be intriguingly close to what
is usually believed to be $\alpha^\prime_\Pom$ of the soft Pomeron.

This observation casts doubts on too straightforward
interpretations of the experimental data on diffraction cone shrinkage 
as a direct measurement of the slope of the Pomeron trajectory.

When studying the $Q^2$-dependence of $\alpha^\prime_{eff}$ within 
the $k_t$-factorization approach, we observed some
decrease of $\alpha^\prime_{eff}$ as we shifted 
from photoproduction limit to DIS. The exact numerical properties
of this decrease depend on the particular choice of the definition of the 
diffraction slope. This should not be forgotten when one compares
the results of the theoretical predictions with the data.
We also analyzed the energy behavior of the compensation effect
and observed that it weakens with energy growth, which agree
with expectations based on the qualitative analysis. Unfortunately, at this
stage we cannot predict the energy behavior of the $\alpha^\prime_{eff}$
beyond the HERA energy range, since such a prediction
requires understanding of the energy behavior of the input parameter
$\alpha^\prime_\Pom$.  

In any case, the mere presence of the dependence of the 
compensation effect on kinematics proves 
that the anti-shrinkage effect is not universal.
This, in turn, supports the understanding that the whole 
underlying picture of the vector meson production 
(the ``real" Pomeron) does not correspond 
to any simple singularity on $j$ plane.\\

I am thankful to Kolya Nikolaev for many valuable comments
and to Igor Akushevich for his help at the early
stage of the code development.
I also wish to thank Prof.~J.Speth for hospitality 
at the Institut f\"ur Kernphysik, Forschungszentrum J\"ulich.
The work was supported by INTAS grants 00-00679 and 00-00366,
and RFBR grant 02-02-17884, and grant ``Universities of Russia" 
UR 02.01.005.


\begin{thebibliography}{99}

\bibitem{VMfirst} B.Z.~Kopeliovich and B.G.~Zakharov, {\em Phys. Rev.} {\bf D44},
3466 (1991); O.~Benhar, B.Z.~Kopeliovich, Ch.~Mariotti,
N.N.~Nikolaev and B.G.~Zakharov,  {\em Phys. Rev. Lett.} {\bf 69}, 1156 (1992).

\bibitem{NNNVM} B.Z.~Kopeliovich, J.~Nemchik, N.N.~Nikolaev, and  B.G.~Zakharov,
{\em Phys. Lett.} {\bf B309}, 179 (1993);
{\em Phys. Lett.} {\bf B324}, 469 (1994).

\bibitem{NNZ94} J.~Nemchik, N.N.~Nikolaev, B.G.~Zakharov, 
{\em Phys. Lett.} {\bf B341}, 228 (1994).

\bibitem{dipole1} N.N.~Nikolaev and B.G.~Zakharov, 
{\em Z. Phys.} {\bf C49} 607 (1991);
{\em Z. Phys.} {\bf C53} 331 (1992);
{\em Z. Phys.} {\bf C64} 631 (1994).

\bibitem{dipole2} A.H.~Mueller, 
{\em Nucl. Phys.} {\bf B415}, 373 (1994);
{\em Nucl. Phys.} {\bf B437}, 107 (1995).

\bibitem{CTreviews} N.N.~Nikolaev, {\em Comments on Nuclear and
Particle Phys.} {\bf 21}, 41 (1992); ``Color transparensy: Novel test of QCD 
in nuclear interactions", {\em Surveys in High Energy Physics}, {\bf 7}, 1 (1994).

\bibitem{VMtalks} I.P.~Ivanov and N.N.~Nikolaev,
``Diffractive vector meson production in $k_t$-factorization approach",
talk given at X International Workshop on Deep 
Inelastic Scattering (DIS2002) Cracow,
Poland, 30 April -- 4 May 2002;
Acta Phys. Polon. {\bf B33}, 3517-3522 (2002);
{\em hep-ph/0206298};
``Diffractive vector meson production in a unified
kappa-factorization  approach'', {\em hep-ph/0006101}, talk given at 8th
International Workshop on Deep Inelastic Scattering and QCD (DIS
2000), Liverpool, England, 25-30 Apr 2000.

\bibitem{disser} I.P.~Ivanov, {\it ``Diffractive vector meson production
in Deep Inelastic Scattering within the $k_t$-factorization approach"},
PhD thesis, 2002, Bonn University, {\em hep-ph/0303053}. 

\bibitem{rhoscale} I.P.~Ivanov, {\em hep-ph/0303063}.

\bibitem{IN2001} I.P.~Ivanov, N.N.~Nikolaev, {\it Phys Rev.}
{\bf D65} 054004 (2002).

\bibitem{NSS} N.N.~Nikolaev, W.~Schafer and G.~Schwiete,
{\em Phys. Rev.} {\bf D63}, 014020 (2001).
[arXiv:hep-ph/0009038].

\bibitem{szczurek} A.~Szczurek, N.N.~Nikolaev, W.~Schafer and J.~Speth,
{\em Phys. Lett.} {\bf B500}, 254 (2001).

\bibitem{BFKL}  V.S.~Fadin, E.A.~Kuraev and L.N.~Lipatov  {\sl Phys. Lett.}
{\bf B60}, 50 (1975); E.A.~Kuraev, L.N.~Lipatov and V.S.~Fadin, 
{\sl JETP} {\bf 44}, 443 (1976); {\bf 45}, 199 (1977);
Ya.Ya.~Balitskii and L.N.~Lipatov, {\sl Sov. J. Nucl. Phys.} {\bf 28}, 
822 (1978).

\bibitem{Lipatov86} L.N.~Lipatov, {\em JETP} {\bf 63}, 904 (1986).

\bibitem{gBFKL} N.N.~Nikolaev, B.G.~Zakharov, and V.R.~Zoller,
{\em JETP Lett.} {\bf 59}, 6 (1994); {\em Phys.Lett.} {\bf B328}, 486 (1994);
{\em J. Exp. Theor. Phys.} {\bf 78}, 806 (1994);
N.N.~Nikolaev and B.G.~Zakharov, {\em J. Exp. Theor. Phys.} {\bf 78}, 
598 (1994); {\em Z. Phys.} {\bf C64}, 631 (1994).

\bibitem{NZZ96}  J.~Nemchik, N.N.~Nikolaev, and B.G.Zakharov,
{\em Phys.Lett.} {\bf B366}, 337 (1996).

\bibitem{NZZ94alphaprime}  N.N.~Nikolaev, B.G.~Zakharov, and V.R.~Zoller,
{\em JETP Lett.} {\bf 60}, 694 (1994).

\bibitem{NNPZZ97} J.~Nemchik, N.N.~Nikolaev, E.Predazzi, B.G.Zakharov and V.R.Zoller,
{\sl J.Exp.Theor.Phys.} {\bf  86} (1998) 1054.

\bibitem{shuvaev} A.G.~Shuvaev, K.J.~Golec-Biernat, A.D.~Martin,
M.G.~Ryskin, {\it Phys.Rev.} {\bf D60}, 014015 (1999).

\bibitem{INS2003} I.P.~Ivanov, N.N.~Nikolaev, A.~Savin, in preparation.

\bibitem{IN99} I.P.~Ivanov, N.N.~Nikolaev, {\it Pis'ma ZhETF (JETP Lett.)}
{\bf 69}, 268 (1999).

\bibitem{ZEUSjpsiphoto} ZEUS Coll., {\it Eur.Phys.J.} {\bf C24}, 345 (2002).

%\bibitem{ZEUSrho} ZEUS Coll., {\it Eur.Phys.J.} {\bf C6}, 603 (1999).

%\bibitem{H1rho} H1 Coll., {\it Eur.Phys.J.} {\bf C13}, 371 (2000).

\bibitem{ZEUSrhophoto} ZEUS Coll., {\it Eur.Phys.J.} {\bf C2}, 247 (1998).

%\bibitem{H1rhophoto} H1 Coll., {\it Eur.Phys.J.} {\bf C13}, 371 (2000).

\bibitem{DL98} A.~Donnachie and P.V.~Landshoff,
{\em Phys. Lett.} {\bf B437}, 408 (1998).

\bibitem{jenk} R.~Fiore, L.L.~Jenkovszky, F.~Paccanoni, and A.~Papa,
{\em Phys.Rev.} {\bf D65}, 077505 (2002); 
R.~Fiore, L.L.~Jenkovszky, F.~Paccanoni, and A.~Papa, {\em hep-ph/0302195}.  

\bibitem{FMS01} L.~Frankfurt, M.~McDermott, and M.~Strikman,
{\em JHEP} {\bf 03}, 045 (2001).

\bibitem{hadronic} F.~Abe et al., {\em Phys. Rev.} {\bf D50}, 5519 (1994).

\bibitem{corrections} P.E.~Volkovitski, A.M.~Lapidus, V.I.~Lisin, 
and K.A.~Ter-Martirosyan, {\em Sov. J. Nucl. Phys.} {\bf 24}, 649 (1976);
A.~Donnachie and P.V.L.~Landshoff, {\em Phys. Lett.} {\bf B123}, 345 (1983);
A.~Capella, J.~Van Trahn and J.~Kwiecinski, {\em Phys. Rev. Lett.}
{\bf 58}, 2015 (1987);
B.Z.~Kopeliovich, N.N.~Nikolaev, and I.K.~Potashnikova, {\em Phys. Rev.}
{\bf D39}, 769 (1989).

\end{thebibliography}
\end{document}